%% file: main.tex
\documentclass[pdftex,twocolumn,epjc3]{svjour3}          

\RequirePackage[T1]{fontenc}

\smartqed  

\RequirePackage{graphicx}
\RequirePackage{flushend}
\RequirePackage[numbers,sort&compress]{natbib}
\RequirePackage[colorlinks,citecolor=blue,urlcolor=blue,linkcolor=blue]{hyperref}
\RequirePackage{subfigure}

\journalname{Eur. Phys. J. C}

\usepackage{rotating}
\usepackage{amsmath}
\usepackage{amssymb}
\usepackage{mathtools}
\usepackage{multirow}
\usepackage{dblfloatfix}
\usepackage{doi}
\usepackage{xspace}
\usepackage{xcolor}
\usepackage{color}
\usepackage{float}
\usepackage{microtype}
\usepackage[switch]{lineno} 
\input{definitions.tex}

\begin{document}
\renewcommand{\arraystretch}{1.3}
\title{Background identification in cryogenic calorimeters through $\alpha-\alpha$ delayed coincidences}

\input{authorlist_2018_BM.tex}
\date{Received:  / Accepted: }
\twocolumn
\maketitle
\begin{abstract}
Localization and modeling of radioactive contaminations is a challenge that ultra-low background experiments are constantly facing.
These are fundamental steps both to extract scientific results and to further reduce the background of the detectors. 
Here we present an innovative technique based on the analysis of $\alpha-\alpha$ delayed coincidences in \th and \u decay chains, developed to investigate the contaminations of the ZnSe crystals in the CUPID-0 experiment. 
This method allows to disentangle surface and bulk contaminations of the detectors relying on the different probability to tag delayed coincidences as function of the $\alpha$ decay position.

\end{abstract}

\section{Introduction}
\label{Sec:Introduction}
\input{1_intro.tex}

\section{Experimental setup}
\label{Sec:Setup}
\input{2_setup.tex}

\section{Data production}
\label{Sec:Processing}
\input{3_dataproduction.tex}

\section{Search for delayed coincidences}
\label{Sec:SearchDelayed}
\input{4_search_delayed.tex}

\section{Localization of crystal contaminations}
\label{Sec:BSAnalysis}
\input{5_analysis_bulk_surf.tex}

\section{Evaluation of delayed coincidence probability}
\label{Sec:MC}
\input{6_monte_carlo.tex}

\section{Results and discussion}
\label{Sec:Results}
\input{7_results.tex}

\section{Conclusions}
\label{Sec:Conclusions}
\input{8_conclusions.tex}

\section{Acknowledgments}
This work was partially supported by the European Research Council (FP7/2007-2013) under contract LUCIFER no. 247115.
We thank M. Iannone for his help in all the stages of the detector assembly, A. Pelosi for constructing the assembly line, M. Guetti for the assistance in the cryogenic operations, R. Gaigher for the mechanics of the calibration system, M. Lindozzi for the cryostat monitoring system, M. Perego for his invaluable help in many tasks, the mechanical workshop of LNGS (E. Tatananni, A. Rotilio, A. Corsi, and B. Romualdi) for the continuous help in the overall set-up design.
We acknowledge the Dark Side Collaboration for the use of the low-radon clean room.
This work makes use of the DIANA data analysis and APOLLO data acquisition software which has been developed by the Cuoricino, CUORE, LUCIFER and \cuz collaborations.
This work makes use of the Arby software for Geant4 based Monte Carlo simulations, that has been developed in the framework of the Milano -- Bicocca R\&D activities and that is maintained by O. Cremonesi and S. Pozzi.
\bibliography{main}  
\bibliographystyle{spphys} 


\end{document}

%% file: definitions.tex
\def\cuz{\mbox{CUPID-0}\xspace}

\def\mone{$\mathcal{M}_{1}$\xspace}
\def\monealfa{$\mathcal{M}_{1\alpha}$\xspace}

\def\mtwo{$\mathcal{M}_{2}$\xspace}

\def\u{$^{238}$U\xspace}

\def\rnddd{$^{222}$Rn\xspace}
\def\poduo{$^{218}$Po\xspace}

\def\th{$^{232}$Th\xspace}
\def\raddq{$^{224}$Ra\xspace}
\def\rnddz{$^{220}$Rn\xspace}
\def\podus{$^{216}$Po\xspace}

\def\onu{$0\nu\beta\beta$\xspace}

\def\Qbb{$Q_{\beta\beta}$}

\def\vita-dim{$T_{1/2}$\xspace}

\def\be{\begin{equation}}
\def\ee{\end{equation}}

%% file: authorlist_2018_BM.tex
\author{O.~Azzolini\thanksref{Legnaro}
\and J.~W.~Beeman\thanksref{LBNL}
\and F.~Bellini\thanksref{Roma,INFNRoma}
\and M.~Beretta\thanksref{MIB,INFNMiB,MattiaPA}
\and M.~Biassoni\thanksref{INFNMiB}
\and C.~Brofferio\thanksref{MIB,INFNMiB}
\and C.~Bucci\thanksref{LNGS}
\and S.~Capelli\thanksref{MIB,INFNMiB}
\and L.~Cardani\thanksref{INFNRoma}
\and P.~Carniti\thanksref{MIB,INFNMiB}
\and N.~Casali\thanksref{INFNRoma}
\and D.~Chiesa\thanksref{MIB,INFNMiB,e1}
\and M.~Clemenza\thanksref{MIB,INFNMiB}
\and O.~Cremonesi\thanksref{INFNMiB}
\and A.~Cruciani\thanksref{INFNRoma}
\and I.~Dafinei\thanksref{INFNRoma}
\and A.~D'Addabbo\thanksref{GSSI,LNGS}
\and S.~Di~Domizio\thanksref{Genova,INFNGenova}
\and F.~Ferroni\thanksref{GSSI,INFNRoma}
\and L.~Gironi\thanksref{MIB,INFNMiB}
\and A.~Giuliani\thanksref{CNRS}
\and P.~Gorla\thanksref{LNGS}
\and C.~Gotti\thanksref{MIB,INFNMiB}
\and G.~Keppel\thanksref{Legnaro}
\and M.~Martinez\thanksref{Roma,INFNRoma,MariaPA}
\and S.~Nagorny\thanksref{LNGS,GSSI,SergePA}
\and M.~Nastasi\thanksref{MIB,INFNMiB}
\and S.~Nisi\thanksref{LNGS}
\and C.~Nones\thanksref{CEA}
\and D.~Orlandi\thanksref{LNGS}
\and L.~Pagnanini\thanksref{LNGS,e2}
\and M.~Pallavicini\thanksref{Genova,INFNGenova}
\and L.~Pattavina\thanksref{LNGS}
\and M.~Pavan\thanksref{MIB,INFNMiB}
\and G.~Pessina\thanksref{INFNMiB}
\and V.~Pettinacci\thanksref{INFNRoma}
\and S.~Pirro\thanksref{LNGS}
\and S.~Pozzi\thanksref{MIB,INFNMiB,e3}
\and E.~Previtali\thanksref{MIB,INFNMiB,LNGS}
\and A.~Puiu\thanksref{GSSI,LNGS}
\and C.~Rusconi\thanksref{LNGS,USC}
\and K.~Sch\"affner\thanksref{GSSI,LNGS}
\and C.~Tomei\thanksref{INFNRoma}
\and M.~Vignati\thanksref{INFNRoma}
\and A.~Zolotarova\thanksref{CNRS}
}
\institute{INFN - Laboratori Nazionali di Legnaro, Legnaro (Padova) I-35020 - Italy \label{Legnaro}
\and
Materials Science Division, Lawrence Berkeley National Laboratory, Berkeley, CA 94720 - USA\label{LBNL}
\and
Dipartimento di Fisica, Sapienza Universit\`{a} di Roma, Roma I-00185 - Italy \label{Roma}
\and
INFN - Sezione di Roma, Roma I-00185 - Italy\label{INFNRoma}
\and
Dipartimento di Fisica, Universit\`{a} di Milano - Bicocca, Milano I-20126 - Italy\label{MIB}
\and
INFN - Sezione di Milano - Bicocca, Milano I-20126 - Italy\label{INFNMiB}
\and
INFN - Laboratori Nazionali del Gran Sasso, Assergi (L'Aquila) I-67100 - Italy\label{LNGS}
\and
Gran Sasso Science Institute, I-67100, L'Aquila - Italy\label{GSSI}
\and
Dipartimento di Fisica, Universit\`{a} di Genova, Genova I-16146 - Italy\label{Genova}
\and
INFN - Sezione di Genova, Genova I-16146 - Italy\label{INFNGenova}
\and
CSNSM, Univ. Paris-Sud, CNRS/IN2P3, Universit\'e Paris-Saclay, F-91405 Orsay - France\label{CNRS}
\and
IRFU, CEA, Universit\'e Paris-Saclay, F-91191 Gif-sur-Yvette, France\label{CEA}
\and
Department of Physics and Astronomy, University of South Carolina, Columbia, SC 29208 - USA\label{USC}
\and
\textit{Present Address}: Physics Department, University of California, Berkeley, CA 94720, USA\label{MattiaPA}
\and
\textit{Present Address}: Fundaci\'on ARAID and Laboratorio de F\'isica Nuclear y Astropart\'iculas, Universidad de Zaragoza, 50009 Zaragoza, Spain\label{MariaPA}
\and
\textit{Present Address}: Queen's University, Physics Department, K7L 3N6, Kingston (ON), Canada\label{SergePA}
}
\thankstext{e1}{davide.chiesa@mib.infn.it}
\thankstext{e2}{lorenzo.pagnanini@lngs.infn.it}
\thankstext{e3}{stefano.pozzi@mib.infn.it}

%% file: 1_intro.tex
\begin{figure*}[b]
\begin{center}
\subfigure{\includegraphics[height=0.24\textwidth]{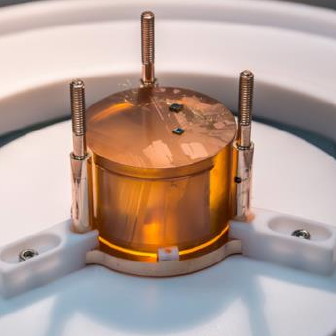}}
\subfigure{\includegraphics[height=0.24\textwidth]{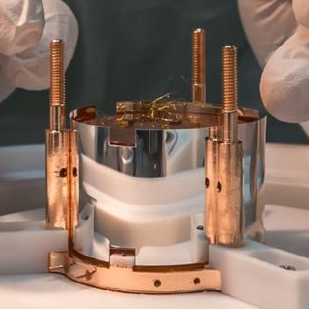}}
\subfigure{\includegraphics[height=0.24\textwidth]{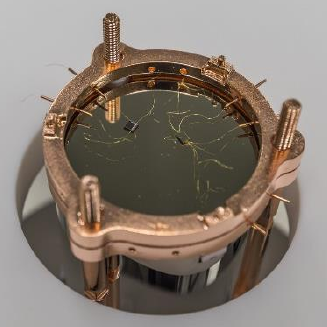}}
\subfigure{\includegraphics[height=0.24\textwidth]{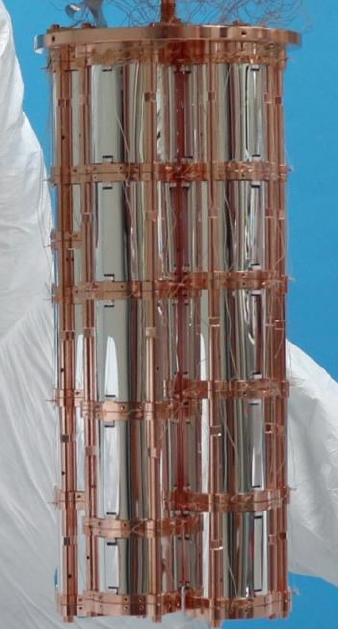}}
\end{center}
\caption{Pictures of the CUPID-0 detector. From left to right: a ZnSe crystal, the same crystal surrounded by the reflecting foil, the Ge light detector mounted on top, the CUPID-0 array of 26 scintillating calorimeters. }
\label{Fig:Detector} 
\end{figure*}

Experiments searching for rare events, such as neutrinoless double-beta (\onu) decay~\cite{Dolinski:2019nrj}, demand for a detailed background understanding in order to implement possible reduction techniques and to collect crucial information for next-generation experiments~\cite{CUPIDInterestGroup:2019inu}.
Depending on the detector features, different techniques are adopted to perform this analysis, exploiting all the information contained in the data themselves, such as particle energy, event topology, time correlation, and particle type.
Cryogenic calorimeters~\cite{Pirro:2017ecr,app11041606} developed to search for the \onu decay have already demonstrated how to use these techniques to understand and reduce the background. 
For example, thanks to the experience gained with the Cuoricino experiment~\cite{Andreotti:2010vj}, a new conceptual design was introduced for the detector holder, thus mitigating the background due to degraded-energy $\alpha$-particles emitted by copper contaminations in the CUORE-0 experiment~\cite{Alessandria:2012zp}. CUORE-0 in turns provided the first background model for cryogenic calorimeters~\cite{Alduino:2016vtd}, on which the CUORE background budget~\cite{Alduino:2017qet} is based.
Over the past decade, we reached a deeper understanding of the background and we further reduced it through scintillating calorimeters~\cite{Pirro:2005ar,Arnaboldi:2010gj,Artusa:2016maw,Armengaud:2017hit}, which introduced the groundbreaking possibility to identify the interacting particles. 

\cuz is the first 10~kg-scale demonstrator of such technique and allowed to reach the lowest background ever measured by cryogenic calorimeters, i.e. $ 3.5 \times 10^{-3}$ counts/ $(\text{keV\,kg\,yr})$ in the region of interest around the $^{82}$Se $\beta\beta$ decay Q-value (\Qbb = 2997.9 $\pm$ 0.3~keV \cite{Lincoln:2012fq}), characterized by an average energy resolution of (20.05 $\pm$ 0.34)~keV FWHM \cite{Azzolini:2019tta,Beretta:2019bmm}.
Such impressive low background rate was achieved by combining the $\alpha$-particle identification (and rejection) with the analysis of time-delayed coincidences. In particular, we tagged potential $^{212}$Bi $\alpha$ decays and we vetoed any event occurring within 7 half-lives of its daughter $^{208}$Tl ($T_{1/2}=3.05$~min), that $\beta$ decays with a high Q-value (5~MeV). In this way, we reduced the background in the region of interest by a factor $\sim$4, at the cost of only 6\% dead time~\cite{Azzolini:2019tta}.

In general, event-tagging based on time-correlations is a widespread tool to identify, quantify, and reduce the background of rare event experiments~\cite{Friedman:2019guf,Hachiya:2020mpu}.
Therefore, we decided to analyze the delayed coincidences due the $\alpha$-decay sequences in \th and \u chains to improve the \cuz background model~\cite{Azzolini:2019nmi}. Indeed, the $\alpha$-decay features, especially the short range of energy deposition, allow to study the contaminant localization.
In this paper, we describe how we analyzed the $\alpha-\alpha$ delayed coincidences in the CUPID-0 data to extract information about the position (bulk vs surface) of crystal contaminations.
This is very important to help designing next-generation bolometric experiments searching e.g. for \onu decay, because the background index induced by such contaminations strongly depends on the their location.

%% file: 2_setup.tex
\cuz is the first 10~kg-scale CUPID \cite{CUPIDInterestGroup:2019inu} demonstrator using enriched scintillating calorimeters to search for \onu decay of $^{82}$Se. 
The \cuz detector is an array of 24 Zn$^{82}$Se crystals 95\% enriched in $^{82}$Se and two ZnSe crystals with natural Se, for a total mass of 10.5~kg. 
When a particle interacts in a ZnSe crystal, it produces a measurable temperature rise proportional to the energy deposit, and a light emission that allows for particle identification.
The typical rise and decay times of signal pulses in ZnSe crystals are 14~ms and 36~ms, respectively~\cite{Azzolini:2018tum}.
The ZnSe crystals are held in a copper frame through small PTFE clamps and laterally surrounded by 70 $\mu$m thick Vikuiti$^{TM}$ reflective foil to enhance light collection. 
To measure the light signal, 170~$\mu$m thick germanium wafers operated as calorimetric detectors~\cite{Beeman:2013zva} are faced to the ZnSe crystals.
Both light detectors and ZnSe crystals are equipped with a Neutron Transmutation Doped (NTD) Ge thermistor \cite{Haller}, acting as temperature-voltage transducer.
The detector is operated at a base temperature of $\sim$10~mK in an Oxford 1000 $^3$He/$^4$He dilution refrigerator located underground in the Hall A of the Laboratori Nazionali del Gran Sasso (Italy). 
The reader can find some pictures of the detector in Fig.~\ref{Fig:Detector} and more details in Ref.~\cite{Azzolini:2018tum}.

%% file: 3_dataproduction.tex
\begin{figure*}[b]
    \centering
    \subfigure{\includegraphics[width = 0.49\textwidth]{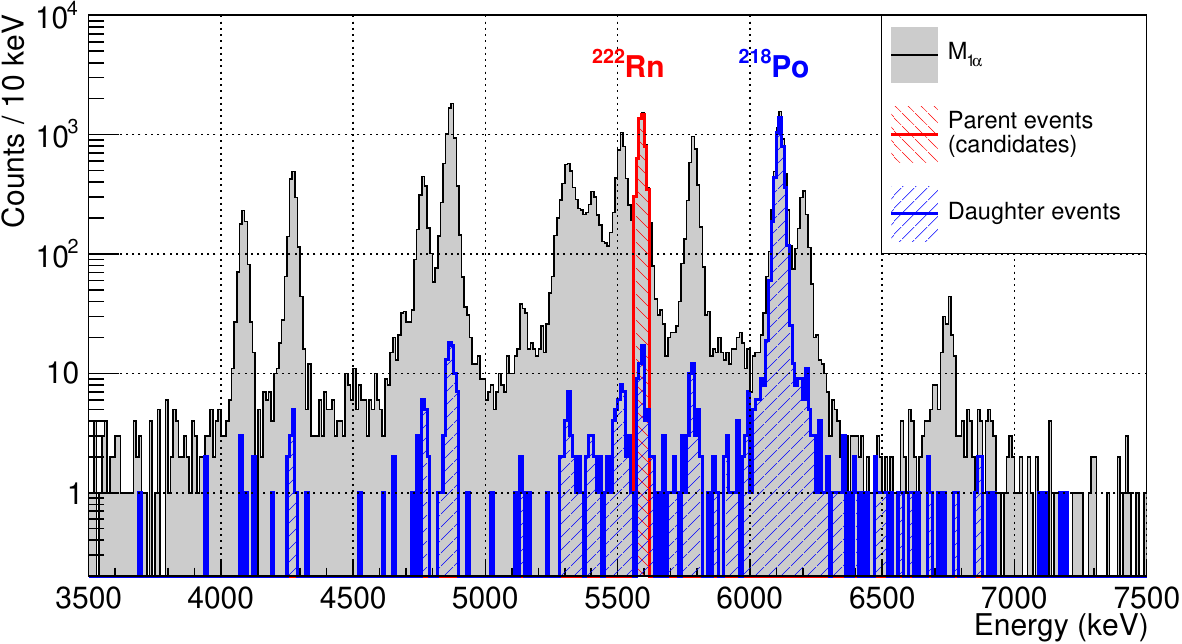}}
    \subfigure{\includegraphics[width = 0.49\textwidth]{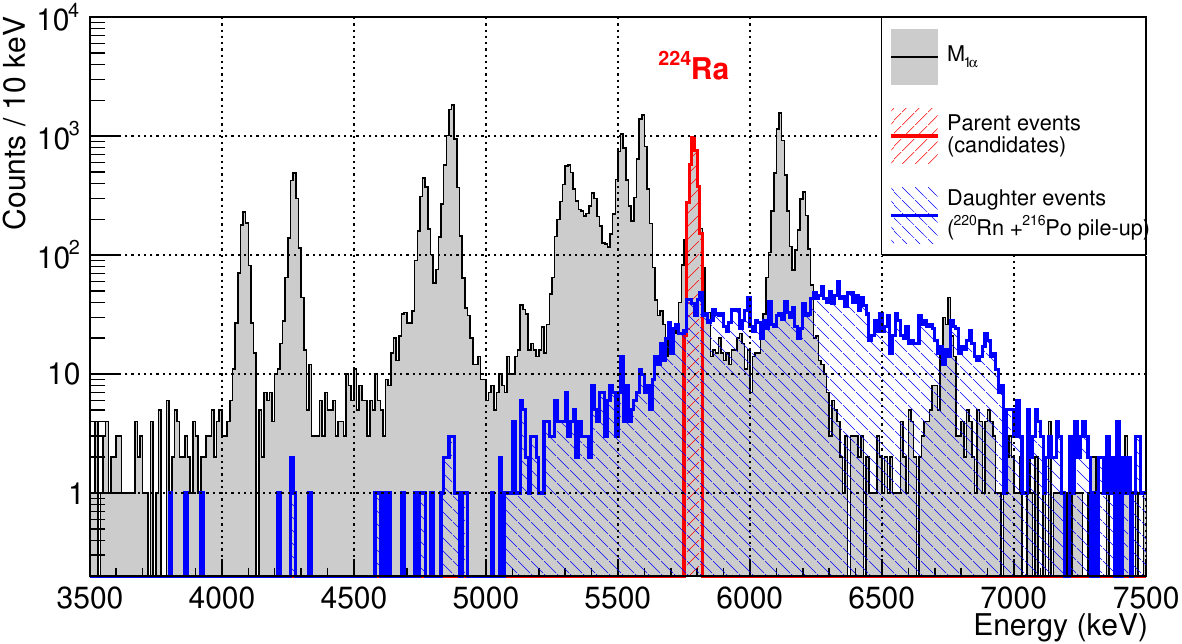}}
    \caption{Search for delayed coincidences in the \u (left) and \th (right) decay chains. The grey spectrum comprises the not piled-up \mone $\alpha$ events. We tag as \textit{daughter} (blue) all the events within a $5 \; T_{1/2}$ time-window after a \textit{candidate parent} event recorded at the $^{222}$Rn (left) or $^{224}$Ra (right) Q-value peak (red). The spectrum of $^{224}$Ra daughters (right, in blue) is miscalibrated due to the $^{220}$Rn$-^{216}$Po pile-up. These events are rejected by the pileup rejection cuts, therefore they are not included in the \monealfa spectrum (grey).}
    \label{fig:DataProduction}
\end{figure*}
In this work, we analyze the spectrum of $\alpha$-particles detected by \cuz Phase I, which lasted from June 2017 to December 2018 with a live-time of 74\% for physics runs.
The continuum data stream from ZnSe detectors is amplified and filtered with a 120 dB/decade, six-pole anti-aliasing active Bessel filter and saved on disk with a sampling frequency of 1~kHz by a custom DAQ software package \cite{DiDomizio:2018ldc}.
We run a derivative trigger to identify the heat pulses and save a 5~s window for each detected signal. 
We save the trigger timestamp of each event with a 1~ms precision, given by the sampling frequency. 
To get the energy deposited in each event, we extract the pulse amplitude by applying a software matched-filter~\cite{Gatti:1986cw}, which improves the signal-to-noise ratio and, thus, the energy resolution.
We convert the pulse amplitude into energy by fitting a parabolic function with zero intercept to the energy of the most intense $\alpha$-peaks produced by \u and \th internal contaminations of ZnSe crystals~\cite{Azzolini:2019nmi}.
We select particle events by requiring a non-zero light signal simultaneously recorded by light detectors and we tag the $\alpha$-particles relying on the light pulse shape parameter defined in Ref.~\cite{Azzolini:2018yye}, which allows to discriminate $>$99.9\% of $\alpha$ events from $\beta/\gamma$ ones at energies $>$2~MeV.
We tag the events that simultaneously trigger more than one crystal within a $\pm$20~ms time window, assigning a \textit{multiplicity} label ($\mathcal{M}_\#$) equal to the number (\#) of crystals hit. Since the total event rate is approximately 50~mHz, the probability of accidental coincidences is almost negligible ($\sim10^{-3}$).
Finally, we analyze the waveform of each triggered event to label piled-up events (1~s before and 4~s after trigger) for which the energy reconstruction is not reliable. 
The data from two enriched crystals, not properly working~\cite{Azzolini:2018tum}, and from the two natural crystals are not considered in the current analysis, therefore the total active mass of the detector is 8.74~kg, with a corresponding exposure of 9.95~kg~yr.

%% file: 4_search_delayed.tex
In \cuz, the $\alpha$-particles are not able to pass through the reflecting foils surrounding the ZnSe detectors, therefore we search for time-correlated events occurring in the same crystal.
The factors that mostly affect the capability to correctly identify delayed coincidences are the time resolution of the detector and the event rate ($r$). The first sets a constraint on the minimum half-life of the daughter nuclide that allows for the two events to be resolved in time. The latter, together with the time window opened to search for delayed coincidences ($\Delta t_w$), determines the probability of random coincidences:

\begin{equation*}
P_\text{random} = 1-e^{-r \Delta t_w} \simeq r \Delta t_w \quad \quad (\Delta t_w \ll 1/r)
\end{equation*}
Since $P_\text{random}$ has to be kept $\ll$1 and $\Delta t_w$ must be chosen of the order of a few half-lives of the daughter nuclide ($T_{1/2}$), the event rate restricts the possibility of searching for delayed coincidences in isotopes with $T_{1/2} \ll 1/r$ only. 

In \cuz, the detector time resolution is of the order of few ms and the rate of $\alpha$ events is at maximum $1.7\times10^{-4}$~Hz/crystal and, on average, $6.3\times10^{-5}$~Hz/crystal.
Therefore, the most suitable $\alpha$-decay sequences in \u and \th chains for this analysis are:

\begin{align*} 
^{222}\text{Rn}&\xRightarrow[3.82~\text{d}]{5.59~\text{MeV}}~^{218}\text{Po}\xRightarrow[186~\text{s}]{6.12~\text{MeV}}~^{214}\text{Pb}
\end{align*}
\begin{align*} 
^{224}\text{Ra}&\xRightarrow[3.66~\text{d}]{5.79~\text{MeV}}~^{220}\text{Rn}\xRightarrow[55.6~\text{s}]{6.4~\text{MeV}}~^{216}\text{Po}\xRightarrow[145~\text{ms}]{6.9~\text{MeV}}~^{212}\text{Pb}
\end{align*}

In order to search for delayed coincidences produced by crystal contaminations, we process the data as follows.
\begin{enumerate}
    \item We label as \textit{candidate parents} ($N_P$) all the single-hit (\mone) not piled-up $\alpha$-events at the Q-value peak of the first decay in the sequence, within a $\pm 1.5 \, \sigma$ energy resolution range.
    This is a good compromise to select a large fraction of candidate parents, without including too much background from the continuum underlying the peaks (see red histograms in Fig.~\ref{fig:DataProduction}). This selection focuses the analysis on crystal contaminations, being the only ones that can produce a signal event at the Q-value. 
    
    \item At each candidate parent, we tag as \textit{daughters} all the events occurring in the same crystal within a time window $\Delta t_w = 5 \, T_{1/2}$ of the second decay (blue histograms in Fig.~\ref{fig:DataProduction}). The length of the coincidence window is optimized to select a large fraction of signal ($\sim$97\%), while keeping random coincidences at a negligible level.
    We only require that a daughter is an $\alpha$-event, without applying multiplicity and pile-up cuts. In this way, we can identify a delayed coincidence even if an uncorrelated event simultaneously triggers another detector or if pile-up occurs. The latter case is particularly frequent in the $^{220}$Rn$-^{216}$Po decay sequence.
    
    \item When two candidate parent events occur in the same detector within $\Delta t_w$, we discard both parents and their daughters from the analysis. In this way, we reduce the contribution from random delayed coincidences and we avoid 
    ambiguity in the assessment of the $\Delta t$ between couples of parent-daughter events. The expected number of these random coincidences between parent candidates is given by:
    
    \begin{equation}\label{eq:randPP}
        N_{PP}=\sum_{ch}N_{P}^{ch} P_\text{random} \simeq \sum_{ch} N_{P}^{ch} r_{P}^{ch} \Delta t_{w}
    \end{equation}
    where $N_{P}^{ch}$ is the number of candidate parent events detected by each channel $ch$, and $r_{P}^{ch}$ is their rate. 
    In Table~\ref{tab:summary} (last row), we check that the number of random coincidences between parent events found in the data ($N^{obs}_{PP}$) is compatible with the expected value calculated through Eq.~\ref{eq:randPP}, finding a very good agreement in both chains.
\end{enumerate}
The energy spectra of parent and daughter events resulting from this analysis are shown in Fig.~\ref{fig:DataProduction}, together with the spectrum of the \mone $\alpha$-events passing the pile-up rejection cut (\monealfa).

In the search for \rnddd$-$\poduo delayed coincidences belonging to the \u chain, the spectrum of daughter events exhibits a clear peak at the \poduo Q-value (Fig.~\ref{fig:DataProduction}~(left)), demonstrating the effectiveness of this technique. 
Since the time window used in this case is relatively long (15.5~min), we also observe random daughter events corresponding to a fraction of $\sim$1\% of the \monealfa spectrum.
To determine the number of detected delayed coincidences ($N_C$), i.e. couples of time-correlated parent-daughter events, we exploit the daughter energy signature and we count the number of events falling in a $\pm 3 \sigma$ energy resolution range centered at the \poduo Q-value.

In the \th chain we have a different situation due to the pile-up between \rnddz and \podus events.
Indeed, when searching for daughters of \raddq decay, most of the selected events are tagged as piled-up and their energies are misreconstructed by the standard data processing (which just discards them). This is why in Fig.~\ref{fig:DataProduction} (right), where we plot all daughter events including those miscalibrated due to the pile-up, we observe a continuous bump instead of two peaks at the \rnddz and \podus Q-values.
Nevertheless, having a good energy reconstruction of these events is not essential, because the information about time correlation is sufficient for the goal of the analysis presented hereafter, which requires $N_C$ to be determined. 
For this purpose, we simply count the number of \raddq events followed by a \rnddz$-$\podus $\alpha$-$\alpha$ piled-up event, that provides an unambiguous signature to identify this sub-chain of 3 consecutive $\alpha$ decays. 
We conservatively discard the piled-up events spaced less than 80~ms in time because above this threshold we are able to precisely trace back the second pulse amplitude to a full-energy \podus decay deposition.

\begin{table}[t]
    \centering
     \caption{Summary of the parameters used to identify delayed coincidences and the corresponding numerical results. 
     As discussed in Sect.~\ref{Sec:BSAnalysis}, the ratio between $N_P$ and $N_C$ depends on the contaminant position. Moreover, in the \th chain, the daughter selection is further constrained to detect 3 consecutive $\alpha$-decays, the third occurring with a $\Delta t>80$~ms.}
    \begin{tabular}{l|cc}
    \hline
     Decay chain     & $^{238}$U  & $^{232}$Th\\
     \hline
     Parent    & $^{222}$Rn & $^{224}$Ra\\
     Daughter  & $^{218}$Po & \rnddz$-$\podus\\
     Parent Range (keV)     & $5590\pm30$& $5789\pm30$ \\
     Daughter Range (keV)   & $6115\pm60$& - \\
     $\Delta t_w$ (s)   & 930  &  278\\
     \hline
     $N_P$ (Parent Candidates)      & 4938 & 3133\\
     $N_{C}$ (Delayed Coincidences)   & 4442 & 2030\\
     $N_{PP}$ / $N^{obs}_{PP}$  & $48 \pm 7$ / 47 &  $6 \pm 2$ / 4\\
     \hline
    \end{tabular}
    \label{tab:summary}
\end{table}

In Table~\ref{tab:summary}, we summarize the parameters used to search for delayed coincidences in \u and \th chains and we report the corresponding results obtained for $N_P$ and $N_C$.
As observed in previous studies~\cite{Pagnanini:2018sgy,Azzolini:2018tum}, contaminations are not homogeneously distributed over all detectors because of an increasing improvement of their radiopurity in the different production batches. Thus, the number of observed delayed coincidences in the different crystals reflects this inhomogeneity. Nevertheless, we find that the $N_C/N_P$ ratio is nearly constant in almost all crystals.

In order to check our selection of delayed coincidences and quantify the amount of random ones, we analyze the time distribution of the $\Delta t$ between couples of parent-daughter events.
Indeed, the $\Delta t$ of physical time-correlated events follows an exponential distribution with a characteristic time parameter equal to the mean-life of the daughter, whereas the $\Delta t$ distribution of random coincidences can be approximated as flat when $\Delta t_w \ll 1/r$.

\begin{figure}
    \centering
    \includegraphics[width = 0.49\textwidth]{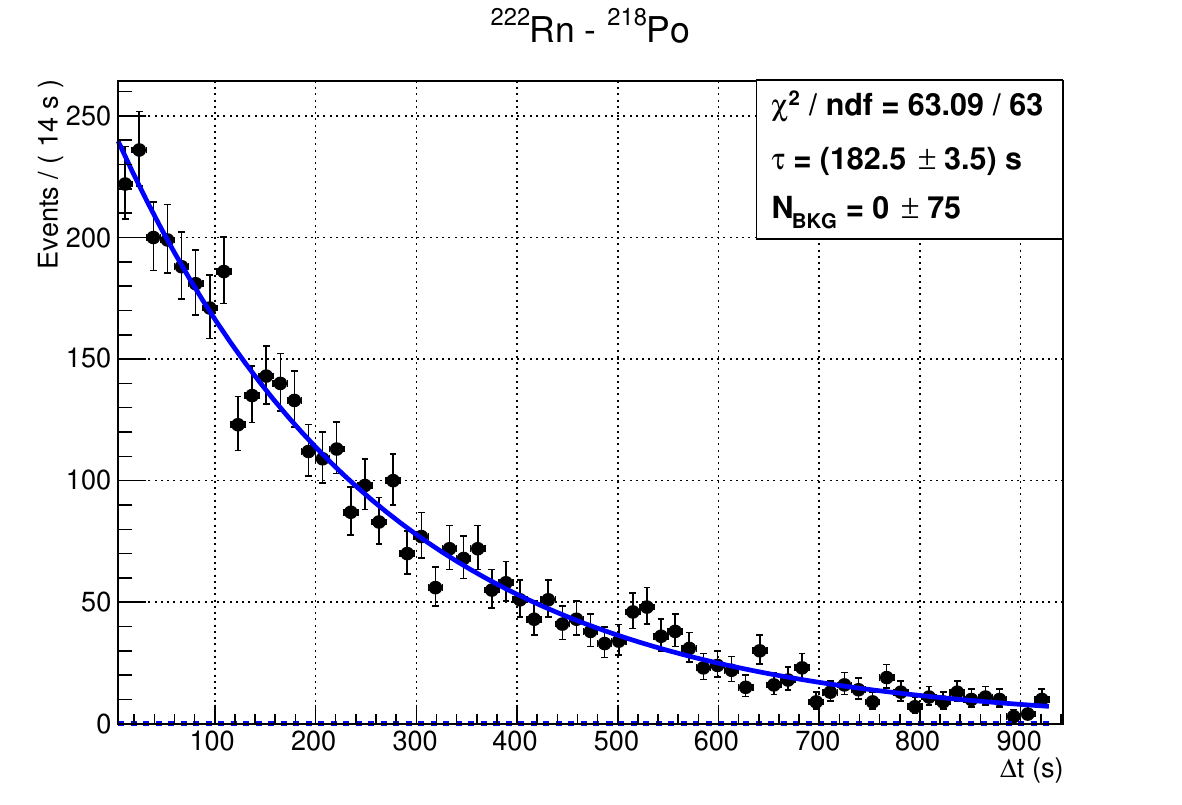}
    \includegraphics[width = 0.49\textwidth]{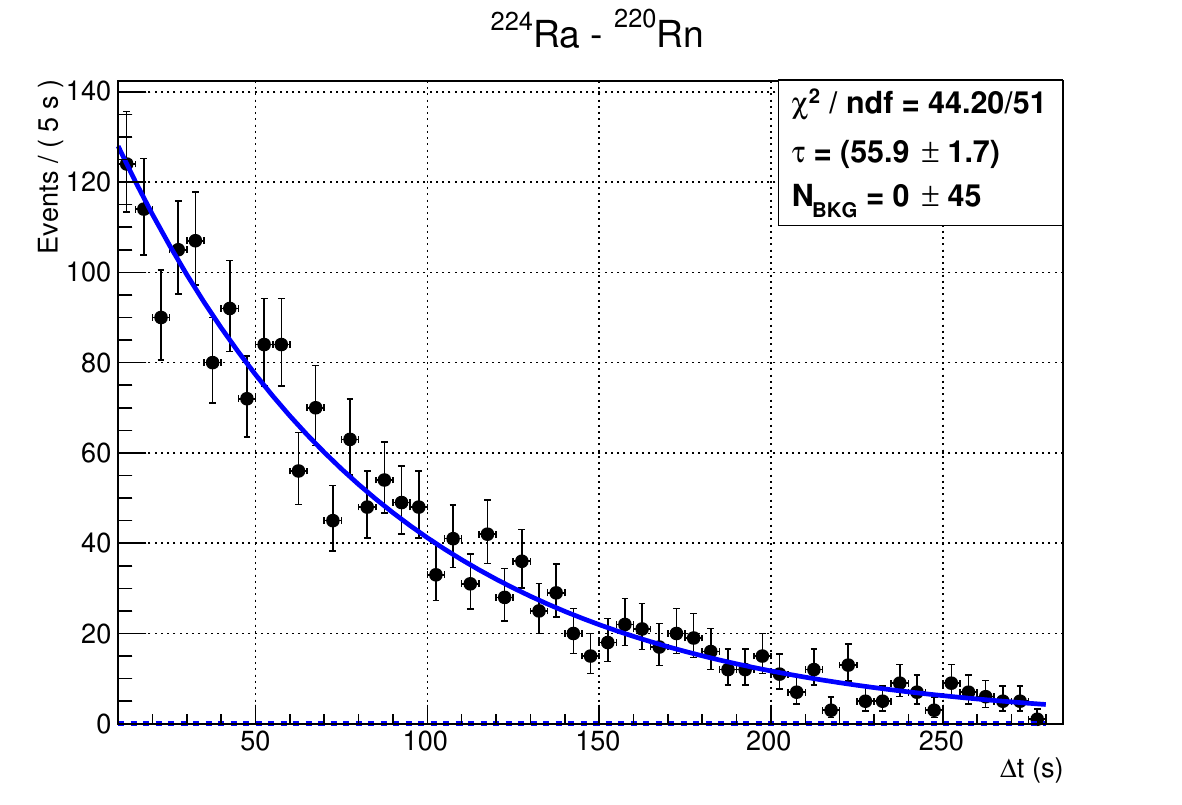}
    \caption{Distribution of the $\Delta t$ between couples of parent-daughter events in the \u ($^{222}$Rn-$^{218}$Po, left) and \th ($^{224}$Ra-$^{220}$Rn, right) decay chains. The fit function (solid line) is composed by an exponential plus a flat background (dashed line at zero counts) to account for possible random delayed coincidences, whose integral ($\text{N}_{\text{BKG}}$) eventually results to be compatible with zero in both cases. The half-life parameter reconstructed by the fit is compatible in both cases with the values reported in literature, thus confirming the effectiveness and the reliability in the identification of delayed coincidences.}
    \label{fig:DeltaTFit}
\end{figure}

As shown in Fig.~\ref{fig:DeltaTFit}, the measured $\Delta t$ of the delayed coincidences identified in the \u (left) and \th (right) chains are distributed with an exponential profile compatible with the half-lives of \poduo~\cite{SINGH2019405} and \rnddz~\cite{BROWNE20111115}, respectively. The flat background results to be compatible with zero in both cases, pointing us out a negligible number of random coincidences.
This is also confirmed by calculating the expected value of random coincidences:

\begin{equation}\label{eq:rnd}
    N_{rnd} \simeq \sum_{ch} (N_P^{ch} - N_{C}^{ch})\, r_{D}^{ch} \Delta t_{w}  
\end{equation}
where $r_{D}^{ch}$ is the $\alpha$-event rate of unpaired daughter-like events (i.e. having the same signature of daughters in term of energy for \poduo or pile-up structure for \rnddz$-$\podus) not in delayed coincidence with a parent. 
According to Eq.~\ref{eq:rnd}, $N_{rnd} \lesssim 1$ in both \u and \th decay chain analysis.

%% file: 5_analysis_bulk_surf.tex
\begin{figure}[t!]
    \centering
    \includegraphics[width=0.5\textwidth]{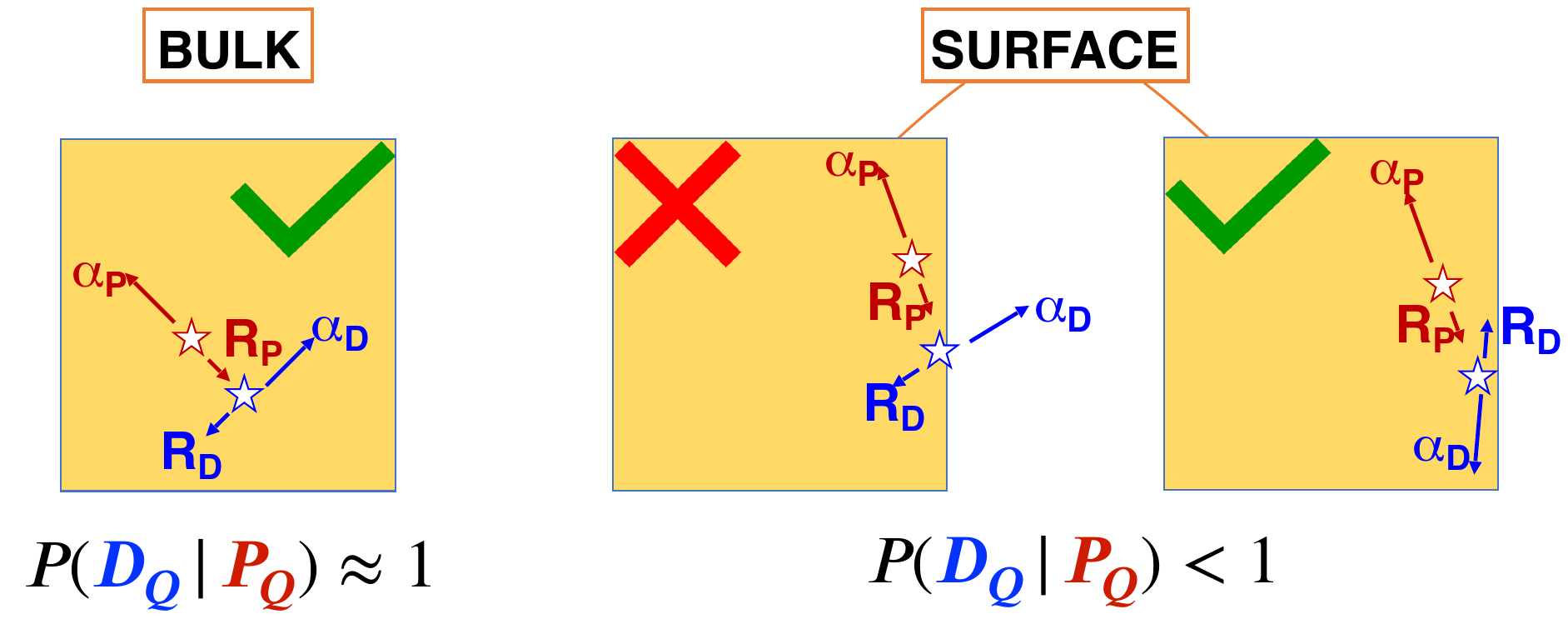}
    \caption{Sketch of $\alpha-\alpha$ delayed coincidences for bulk (left) and surface (right) crystal contaminations. Parent and daughter decays are represented in red and blue, respectively. In the bulk case, there is nearly a 1:1 ratio between detected daughter events and parent candidates. This ratio falls below 1, when contaminations are on the detector surfaces due to the escape of the $\alpha$ emitted in the daughter decay.}
    \label{fig:surfvsbulk}
\end{figure}

In the experiments searching for rare events, it is fundamental to localize the radioactive contaminations because the background rejection techniques have different efficiencies depending on their position.
In cryogenic calorimeters, surface contaminations are of particular concern because the whole crystal volume is sensitive in detecting particle interaction, without any dead-layer.
A very effective way traditionally used to identify surface contaminations of cryogenic calorimeters consists in analyzing the spectrum of \mtwo events comprised of $\alpha-$recoil coincidences in neighbours crystals~\cite{Alduino:2016vtd}.
This method cannot be applied to CUPID-0 Phase I data analysis, because the reflective foil around the ZnSe crystals absorbs the $\alpha$-particles escaping from their surfaces, preventing the detection of $\alpha-$recoil \mtwo events.
To overcome such limitation, we conceived an innovative method based on the analysis of $\alpha$-$\alpha$ delayed coincidences. As shown in the next section, both bulk and surface contaminations produce candidate parent events at the Q-value, allowing to search for delayed coincidences with the procedure introduced in Sect.~\ref{Sec:SearchDelayed}.
Since the ratio between the number of detected delayed coincidences ($N_C$) to the number of candidate parents ($N_P$) depends on the source location, we can extract information about it.
As sketched in Fig.~\ref{fig:surfvsbulk}, if the contamination is in the crystal bulk, the probability to detect a full-energy daughter event given a candidate parent observed at the Q-value, $p \left( D_Q|P_Q \right)$, is almost 1.
Conversely, when contaminations are on crystal surfaces, the $\alpha$-particles have a not negligible probability to escape the detector. 
Thus in this case, the conditional probability $p \left( D_Q|P_Q \right)$ to detect a daughter event at the Q-value is significantly $<1$.

In order to determine the activity ratio $r = A_s / A_b$ between surface ($s$) and bulk ($b$) contaminations of a particular decay sub-chain, we solve the following system of equations:

\begin{equation}
\begin{cases}
    N_P = A_b \, T (\varepsilon_P^b + r \, \varepsilon_P^s)\\
    N_C = A_b \, T (\varepsilon_C^b + r \, \varepsilon_C^s) 
\end{cases}
\label{Eq:system}
\end{equation}
in which the number of \textit{candidate parents} ($N_P$) and the number of \textit{delayed coincidences} ($N_C$) are expressed as a function of the contaminant activities ($A$), the measurement livetime ($T$), and the detection efficiencies ($\varepsilon$).
In this system, the different value of $p \left( D_Q|P_Q \right)$ exploited to disentangle bulk and surface contaminations (hereafter labeled as $p_C$) enters in the $\varepsilon_C$ terms, which can be expressed as 
$\varepsilon_C = \varepsilon_P \, p_C$.

By solving the system in Eq.~\ref{Eq:system}, we finally obtain the formula to calculate $r$:

\begin{equation}
r = \frac{\varepsilon_P^b \left(N_P \, p_C^b - N_C \right)}{\varepsilon_P^s \left(N_C - N_P \, p_C^s \right)}
\label{Eq:solution}
\end{equation}

The physical constraint to be respected in order to get positive results for $r$ is:

\begin{equation}
    p_C^s \leq N_C / N_P \leq p_C^b
\end{equation}
This is consistent with the fact that in an experiment we can expect to observe delayed coincidences from a combination of bulk and surface contaminations. If one of them is dominant, the experimental ratio $N_C / N_P$ will approach the range limits.

%% file: 6_monte_carlo.tex
\begin{figure*}[t]
\begin{center}
\includegraphics[width=\textwidth]{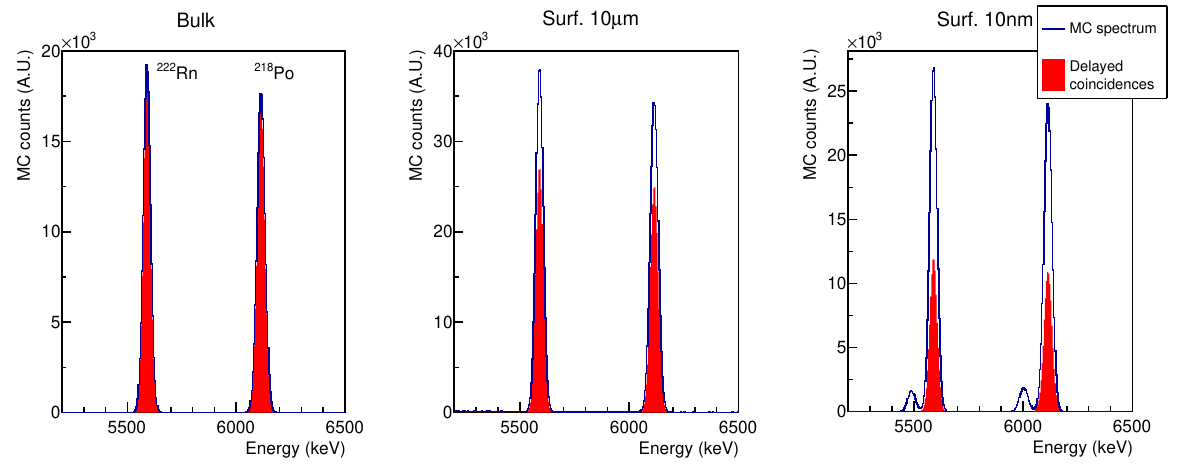}
\end{center}
\caption{Monte Carlo spectra of \u decay chain, zoomed on \rnddd and \poduo peaks, obtained by simulating the contaminants in the crystal bulk (left), and on crystal surfaces with depth parameters of 10~$\mu$m (center) and 10~nm (right). The fraction of \rnddd--\poduo delayed-coincidences (highlighted in red) over the total events recorded at the Q-value peaks decreases as the contaminants are simulated in a thinner surface layer due to $\alpha$-particle escapes. The small peaks appearing in the rightmost plot are due to recoil escapes occurring when contaminants are simulated in a very shallow surface layer ($\lambda=10$~nm).}
\label{Fig:MC} 
\end{figure*}

In the previous section, we showed that the ratio between the activity of surface and bulk crystal contaminations can be determined from the experimental data once the efficiencies ($\varepsilon_P^b$ and $\varepsilon_P^s$) and the probabilities to detect delayed coincidences ($p_C^b$ and $p_C^s$) are known. 
We evaluate these parameters through Monte Carlo simulations.

We simulate the background sources identified in Sect.~\ref{Sec:SearchDelayed} with a Monte Carlo tool, named \textit{Arby}, based on the Geant4 toolkit~\cite{Geant4}, version 10.02. 
The particles generated by the radioactive decays of interest are propagated using the G4EmLivermore physics list. 
The decay chains of \th and \u can be simulated completely or in part, to reproduce secular equilibrium breaks.
For each energy deposit in the detector, we record the information about the crystal where the interaction took place, the amount of deposited energy, and the time elapsed since the previous event in the decay chain. 
In order to make the simulation output as similar as possible to the experimental data, we implement the detector response and the data production features in a second step. 
We associate to each decay an absolute time randomly sampled on the time scale of the experimental data taking, with the exception of the decays occurring within 1 hour from their predecessors in order to preserve the time correlations of interest for this analysis.
We account for the detector time resolution by summing up the energy depositions that occur in the same crystal within a $\pm5$~ms window, and we group into multiplets the events involving different crystals within $\pm20$~ms.

We simulate bulk contaminations by generating the decays in random positions uniformly distributed within the whole crystal volume.
The usual approach for simulating surface contaminations in cryogenic calorimeters is to sample the decay positions from an exponential distribution $e^{-x/\lambda}$, where $\lambda$ is the so-called depth parameter. 
This parameter is usually chosen in the range from few nm up to tens of $\mu$m, in order to reproduce the signatures from shallower to deeper contaminations observed in the experimental data~\cite{Alduino:2016vtd, Azzolini:2019nmi}. 
Different values of the depth parameter can be traced back to different contamination mechanisms. For example, the exposure of a material to the air is expected to produce a very shallow contamination, whereas the treatment of crystal surfaces~\cite{ARNABOLDI20102999} can originate deeper contaminations.

In Fig.~\ref{Fig:MC} we show the result obtained for the \u chain simulated in the bulk of crystals and on their surfaces with two different depth parameters. 
We set the two $\lambda$ values equal to 10~nm and 10~$\mu$m, being much lower and on the same scale of the $\alpha$-particle range, respectively.
Even for very shallow surface contaminations, we get a significant fraction of events reconstructed at the $\alpha$-decay Q-value.
We process the Monte Carlo outputs with the same procedure used to tag the delayed coincidences in the experimental data and we highlight the parent-daughter coincident events in the plot. 
As expected, the ratio of delayed coincidences over the number of parent candidates decreases as the contamination is simulated in a shallower layer near the crystal surfaces.

The analysis method presented in Sect.~\ref{Sec:BSAnalysis}, provides a single parameter to quantify the activity of surface contaminations, thus a unique model must be chosen to describe them.
According to Fig.~\ref{Fig:MC}, the deeper surface contaminations (10~$\mu$m) produce a delayed coincidence signal which can be viewed as a combination of a bulk contamination and a shallower surface one. 
Therefore, in our analysis, we choose the simulations with $\lambda=10$~nm to model surface contaminations and to study the ratio between bulk and surface activities. 

\begin{table}[]
    \centering
    \caption{Detection efficiencies of candidate parents ($\varepsilon_P$) and probabilities of delayed coincidences ($p_C$) evaluated from Monte Carlo simulations of bulk ($b$) and surface ($s$) crystal contaminations. For the surface contaminations, we sample the decay positions from an exponential distribution with $\lambda = 10$~nm. Uncertainties are negligible due to the high Monte Carlo statistics.}
    \begin{tabular}{c|cc}
    Isotope & Bulk & Surface\\
    \hline
    \rnddd & $\varepsilon_P^b$ = 84.9\% & $\varepsilon_P^s$ = 39.6\%\\
    \poduo & $p_C^b$ = 96.5\% & $p_C^s$ = 44.2\% \\ 
    \hline
    \raddq & $\varepsilon_P^b$ = 79.5\% & $\varepsilon_P^s$ = 36.1\%\\
    \rnddz -- \podus & $p_C^b$ = 65.7\% & $p_C^s$ = 14.3\% \\ 
    \hline
    \end{tabular}
    \label{tab:MC}
\end{table}

In Table~\ref{tab:MC}, we report the values of $\varepsilon_P^b$, $\varepsilon_P^s$, $p_C^b$, and $p_C^s$ obtained from the MC simulations of \u and \th decay chains.
The $\varepsilon_P$ efficiencies are computed by taking into account that a $\pm 1.5 \, \sigma$ range was used to select the candidate parents in the experimental data.
As expected from a simple geometric reasoning about $\alpha$ escape process, the efficiencies and the probabilities related to surface contaminations are about half with respect to the bulk ones.
The only exception is the probability to detect a delayed \rnddz--\podus piled-up event. This is because we are searching for a triple $\alpha$-decay sequence and we have to discard a fraction of piled-up events with $\Delta t<80$~ms to be consistent with the experimental data processing.

%% file: 7_results.tex
\begin{table}[b]
    \centering
    \caption{Experimental input and final results of the delayed coincidence analysis. Both for \u and \th decay chain we quote: the number of parent candidates ($N_P$), net of background subtraction; the number of detected coincidences ($N_C$) with their binomial uncertainties; the ratio between surface and bulk contamination activities ($r$) and their absolute values ($A_b$, $A_s$). We quote the $r$ result for \th chain with an asymmetric uncertainty range to exclude negative non-physical values.}
    \begin{tabular}{c|c|c}
    Decay chain & \u & \th\\
    \hline
    $N_P$ & $4868 \pm 70$ & $3118 \pm 56$\\
    $N_C$ & $ 4442 \pm 20$ & $ 2030 \pm 27$\\
    $N_C / N_P$ & $ (91.2 \pm 0.4) \%$ & $(65.1 \pm 0.9) \%$ \\
    $r$ & 0.24 $\pm$ 0.06 & $0.03^{+ 0.11}_{-0.03}$\\
    $A_b$ ($\mu$Bq)  & $146 \pm 4$ & $108 \pm 5$\\
    $A_s$ ($\mu$Bq)  & $35 \pm 8$ & $3^{+ 11}_{-3}$\\
    \end{tabular}
    \label{tab:results}
\end{table}

In this section we report the results of the delayed coincidence analysis based on the \cuz data, obtained by combining the experimental data ($N_P$ and $N_C$) with the Monte Carlo evaluations summarized in Table~\ref{tab:MC}.
The $N_P$ values reported in Table~\ref{tab:summary} can not be directly used to calculate the activity ratio $r$, because they include a fraction of background events. 
Indeed, other radioactive sources can produce some events falling in the energy range of candidate parents.
We exploit the \cuz background model~\cite{Azzolini:2019nmi} to assess such contribution, which results on the percent scale for both parent peaks in the \th and \u chains.
The $N_P$ values obtained after subtracting the expected background counts are reported in Table~\ref{tab:results}, with an uncertainty that takes into account the Poisson fluctuations.
The uncertainty associated to $N_C$ is the Binomial one with $N_C$ successes given $N_P$ trials.
After calculating $r$ with Eq.~\ref{Eq:solution}, we solve the system in Eq.~\ref{Eq:system} to get $A_s$ and $A_b$. 

The results of this analysis prove that most of \cuz crystal contaminants are located in their bulk. 
For the \u sub-chain we get that $\sim$20\% of decays occur near crystal surfaces, whereas for the \th sub-chain this fraction is constrained in a range between zero and a 13\% 1~$\sigma$ upper limit.
This information was used to set prior constraints in the \cuz background model~\cite{Azzolini:2019nmi}, allowing for the disentanglement of surface vs bulk crystal contaminations.

Given the total mass ($m=8.74$~kg) and surface ($S=2149$~cm$^{2}$) of ZnSe crystals used for this analysis, we calculate the specific activities of the $\alpha$-decay sequences for the \u sub-chain:

\begin{align*}
A_b / m &= (16.7 \pm 0.5) \; \mu\text{Bq/kg} \\
A_s / S &= (16 \pm 4)  \; \text{nBq/cm}^2
\end{align*}
and for the \th one:

\begin{align*}
A_b / m &= (12.4 \pm 0.6) \; \mu\text{Bq/kg} \\
A_s / S &= (1.4 ^{+ 5}_{-1.4}) \; \text{nBq/cm}^2
\end{align*}

It is worth noting that, because of secular equilibrium break, these results refer only to the second parts of \u and \th decay chains, which are characterized by higher activities with respect to the first parts (see \cite{Azzolini:2019nmi} for more details).

\subsection{Systematics discussion}
The results of this analysis depend on the efficiencies and probabilities related to surface contaminations reported in Table~\ref{tab:MC}. 
These parameters are affected by the escape probabilities of $\alpha$s and nuclear recoils.
The $\alpha$ escape probability is significantly affected when the contamination depth is changed from $\lambda=10$~nm to a value on the same scale of the $\alpha$ range. For example, by analyzing our data with $\lambda=10$~$\mu$m, the activity ratios $r$ would scale up by a factor $\sim 2$. This confirms that, for our analysis, a deep surface contamination is equivalent to a combination of a bulk and a shallow surface contamination.
On the other hand, according to our MC simulations, we can consider the nuclear recoil escape as a second order effect for $\lambda \gtrsim 10$~nm. Since in the experimental spectrum there are no emerging peaks at the $\alpha$ energies of the isotopes analyzed in this work, we can exclude that shallower contaminations ($\lambda\ll$10~nm) can significantly affect our results.

%% file: 8_conclusions.tex
In this work, we presented an innovative analysis technique to study the background sources in cryogenic calorimeters relying on the time-correlation of $\alpha$-decay sequences in \u and \th chains. 
This method allowed us to disentangle surface and bulk contaminations of ZnSe crystals exploiting the different probability to detect delayed coincidences depending on the contamination depth (see Fig.~\ref{Fig:MC}). 
In particular, we demonstrated that the \u and \th contaminants of \cuz detectors are mainly located in the bulk of crystals.
This technique, that was applied for the first time to set prior constraints in the \cuz background model~\cite{Azzolini:2019nmi}, can be adopted also in other experiments for broader purposes.
For example, in the analysis of CUORE data~\cite{Adams:2021rbc}, delayed coincidences could help to better constrain the background sources~\cite{Adams:2021xiz} and to reject the time-correlated events falling in the region of interest.
Moreover, the R\&D activities for CUPID~\cite{Armatol:2020oxe,Armatol:2020rtb}, CUPID-Mo~\cite{Armengaud:2019rll,Armengaud:2019loe,Armengaud:2020luj}, and in general the experiments searching for rare events can profit from this technique to study the radioactive contaminations of detector components and to select ultra-pure materials, with also the possibility to analyze other decay sequences in \u, \th and $^{235}$U chains~\cite{Nastasi:2021}. 
Finally, the analysis of delayed coincidences in \cuz allowed to measure the half-life of \podus~\cite{Beretta:2021}.